\documentclass{WileyMSP-template}

\usepackage{chemformula} 
\usepackage[T1]{fontenc} 
\usepackage{url}
\usepackage{siunitx}    
\usepackage{float}      
\usepackage[%
    pdfborder={0 0 0}
]{hyperref}
\usepackage{parskip}
\usepackage{microtype}
\usepackage[superscript]{cite}
\usepackage[final]{changes}

\makeatletter
\renewcommand{\@citess}[1]{\textsuperscript{[#1]}}
\renewcommand{\@biblabel}[1]{#1.} 
\makeatother

\usepackage{caption}
\usepackage{titlesec}
\titlelabel{\thetitle.\quad}

\titleformat{\subsection}
  {\large\itshape}{\thesubsection.}{1em}{}
\titleformat{\subsubsection}
  {\large\itshape}{\thesubsubsection.}{1em}{}

\begin{document}

\title{Optical Signature of Moiré Superlattices Formed by Twisted $\text{SrTiO}_3$ Membranes}
\maketitle

\author{T. A. M. Ragib Shahriar${^\dag}$}
\author{Fumikazu Murakami${^\dag}$}
\author{Xing He${^\dag}$}
\author{Konnor Koons}
\author{Xinyan Li}
\author{Bumseop Kim}
\author{Shihan Qin}
\author{Varun Harbola}
\author{Jochen Mannhart}
\author{Yimo Han}
\author{Ruijuan Xu${^*}$}
\author{Shengxi Huang${^*}$}
\author{Andrew Rappe${^*}$}
\author{Hanyu Zhu${^*}$}

\begin{affiliations}
T. A. M. Ragib Shahriar, Dr. Xinyan Li, Prof. Yimo Han, Prof. Hanyu Zhu\\
Materials Science and NanoEngineering, Rice University, Houston, United States\\
Smalley Curl Institute, Rice University, Houston, United States\\
Rice Advanced Materials Institute, Rice University, Houston, United States\\
E-mail: hanyu.zhu@rice.edu

Dr. Fumikazu Murakami, Prof. Shengxi Huang\\
Electrical and Computer Engineering, Rice University, Houston, United States\\
Smalley Curl Institute, Rice University, Houston, United States\\
Rice Advanced Materials Institute, \added{Rice University,} Houston, United States\\
E-mail: shengxi.huang@rice.edu

Dr. Xing He, Dr. Bumseop Kim, Shihan Qin, Prof. Andrew M. Rappe\\
Department of Chemistry, University of Pennsylvania, Philadelphia, United States\\
E-mail: rappe@sas.upenn.edu

Konnor Koons, Prof. Ruijuan Xu\\
Materials Science and Engineering, North Carolina State University, Raleigh, United States\\
E-mail: rxu22@ncsu.edu

Dr. Varun Harbola{$^\ddagger$}, Prof. Jochen Mannhart\\
Max Planck Institute for Solid State Research, Stuttgart, Germany

\deleted{Corresponding to: hanyu.zhu@rice.edu, rappe@sas.upenn.edu, shengxi.huang@rice.edu, rxu22@ncsu.edu}

\added{Funding: T.A.M.R.S. and H.Z. acknowledge support from the National Science Foundation (NSF) under grant number DMR-2240106 and the Welch Foundation (C-2311). X.L. acknowledges support from the Rice Advanced Materials Institute (RAMI) at Rice University as a RAMI Postdoctoral Fellow. X.L. and Y.H. acknowledge support from the NSF (FUSE-2329111 and CMMI-2239545) and the Welch Foundation (C-2065). K.K. acknowledges support from the Army Research Office under award No. W911NF-25-1-0201. R.X. acknowledges support from the NSF under award No. DMR-2442399. F.M. and S.H. acknowledge support from the NSF (FUSE-2329111, ECCS-2246564, and ECCS-1943895) and the Welch Foundation (C-2144). B.K. and A.M.R. acknowledge support from the U.S. Department of Energy, Office of Science, Basic Energy Sciences, under Award No. DE-SC0026196 for analysis of moiré band structure and optical spectroscopy. X.H., S.Q., and A.M.R. acknowledge support by the Office of Naval Research under grant number N00014-24-1-2500 for the study of bonding, composition, and structural dynamics in SrTiO$_3$ interfaces.}

\end{affiliations}








\begin{abstract}
Moiré superlattices formed at the interfaces of mismatched lattices have attracted significant interest over the past decade due to their large tunability of band parameters and interactions among electrons, spins, and lattices. Superlattices made from twisted perovskite oxides may have strong structure and potential modulation, but evidence of such modulation over macroscopic areas, particularly at large twisting angles, has not been clearly demonstrated so far. Here, \replaced{millimeter-scale twisted oxide bilayers are fabricated}{we fabricated millimeter-scale twisted oxide bilayers} at $36^\circ$ angle, close to the simple coincidence site lattice condition $\Sigma5$, from freestanding SrTiO$_3$ membranes. We \replaced{observe}{discovered} new low-frequency vibrational modes whose Raman activity, according to molecular dynamics simulations, is greatly enhanced by an asymmetric, twisted interface between the SrO and TiO$_2$ layers. Such an interface is energetically favorable from first-principles calculations and is corroborated by the observation of strong second\added{-}harmonic generation from the interface comparable to that from the SrTiO$_3$ surface throughout the bilayer region. The results are consistent with interlayer coupling enhanced by high-temperature annealing and confirmed by cross-sectional scanning transmission electron microscopy imaging. Our work sheds light on the structural behavior of twisted oxides and provides directions for tuning their phononic and nonlinear optical properties in future studies.
\end{abstract}

{$^\ddagger$}Present address: Department of Condensed Matter Physics \& Materials Science, Tata Institute of Fundamental Research, Mumbai 400005, India

${^\dag}$These authors contributed equally to this work.

\section{Introduction}
Moiré superlattices created by stacking two periodic lattices with a small relative twist or lattice mismatch have emerged as a powerful route to engineer electronic structure, correlations, and lattice textures beyond what is possible in bulk crystals or conventional heterostructures~\cite{andrei_marvels_2021}. A variety of exotic quantum phenomena can be obtained through moiré engineering, including unconventional superconductivity, correlated and topological states, orbital magnetism, etc., with great potential to advance fundamental science and develop new device applications in electronics and photonics~\cite{kennes_moire_2021, lau_reproducibility_2022,cai_signatures_2023}. The field of twisted moiré heterostructures started with van der Waals materials, but recent efforts have explored the possibility of moiré physics in a broader range of material systems, including complex perovskite oxides, where the additional lattice, orbital, and polar degrees of freedom offer an even richer phase space for emergent moiré phenomena~\cite{li2022stacking, Pryds_2024}. Distinct from van der Waals moiré systems, the strong ionic-covalent interlayer bonding in perovskite oxides potentially leads to robust structural and electronic coupling across moiré interfaces, offering a promising platform for exploring a richer landscape of emergent moiré phenomena. These characteristics have driven rapidly growing research interest, with theoretical predictions of intriguing phenomena like robust ferroelectricity, moiré-periodic charge modulation, topological Lieb lattices, flat bands, correlated electronic states, and unconventional magnetic states~\cite{lee_moire_2024, xu_creating_2025, shahed_prediction_2025}. The experimental demonstration of oxide moiré systems is enabled by advances in the fabrication of crystalline oxide membranes with millimeter lateral dimensions and precisely controlled thicknesses down to the unit-cell level~\cite{lu_synthesis_2016,xu_strain-induced_2020}. For example, membrane bilayers assembled at a controlled twist angle exhibit flexoelectricity-driven polar vortex-antivortex networks~\cite{sanchez-santolino_2d_2024}, as well as interfacial electronic states that vary periodically within the moiré supercell according to the local atomic registry~\cite{kim_charge_2025}. 

However, unlike van der Waals materials, perovskite oxides usually exhibit sensitive surfaces whose properties depend on the types of termination, which have long been a challenge for forming atomically coherent interfaces in oxide heterostructures, even between lattice-matched materials~\cite{zubko_interface_2011}. In addition, the surface roughness of the membranes hinders interfacial coupling over extended lateral areas, limiting atomic-scale proximity and coherent bonding at the interface.~\cite{Kp2025-he}. Recently, the formation of clean and sharp interfaces was shown between symmetry-incompatible cubic SrTiO$_3$ membranes and hexagonal sapphire substrates via CO$_2$ laser annealing at temperatures exceeding 1000 $^{\circ}$C, highlighting the potential of high-temperature annealing to achieve atomically clean interfaces in oxide moiré systems~\cite{wang_interface_2024}. Yet, to date, experimental evidence from electron microscopy only shows changes in local structural and charge states~\cite{kim_charge_2025}. Possible long-range symmetry-breaking properties originating from the moiré superlattices, such as the electronic and vibrational wavefunctions, have not been investigated optically like in van der Waals 2D systems~\cite{huang_low-frequency_2016,lin_moire_2018,lin_large-scale_2021,quan_phonon_2021,quan_quantifying_2023,ci_breaking_2022,paradisanos_second_2022,du_nonlinear_2024,zhu_creating_2024}. Moreover, although the surface of perovskites has been studied in depth,\cite{padilla1998ab,kawasaki1994atomic,erdman2002structure} the interface structure of the twisted oxide membranes is still under debate from the theoretical perspective~\cite{xu_creating_2025,lee_moire_2024,kim_charge_2025}. In perovskite oxides, varying the interfacial terminations can lead to specific stacking registries, polarization, and structural distortions, thereby altering the energy landscape and the symmetry of the moiré bilayers. For SrTiO$_3$, recent works postulated different interface structures (SrO-SrO, TiO$_2$-TiO$_2$, and SrO-TiO$_2$ types), which should exhibit different thermodynamic stability across the chemical potential phase diagram and lead to distinct functionalities.\cite{lee_moire_2024,kim_charge_2025} Therefore, a comprehensive understanding of the interfacial structures of moiré superlattices from first principles is important since the interface itself can affect the functionality by different bond formation, symmetry breaking, electronic reconstruction, and lattice dynamics.

Here we report the successful fabrication of a twisted SrTiO$_3$ bilayer with a 36$\degree$ [001] twist, close to the simple coincident site lattice condition $\Sigma=5$, which shows unique vibrational and nonlinear optical responses. The bonded interface was formed by stacking two freestanding SrTiO$_3$ membranes, followed by high-temperature laser-annealing. Using macroscopic, non-destructive optical probes such as Raman scattering and second-harmonic generation, we directly investigated the interfacial phonon modes and electronic symmetry-breaking arising from the interfacial atomic reconstruction. A low-frequency Raman mode appears in the bilayer at temperatures both above and below the tetragonal phase transition of bulk SrTiO$_3$, indicating symmetry breaking that is independent of the bulk lattice distortion. Using \textit{ab initio} molecular dynamics (AIMD) simulation, we found that the brightening of the interfacial shear mode may be attributed to an asymmetric bonding at the SrO-TiO$_2$ interface. The alternating layers of SrO and TiO$_2$ are thermodynamically favored (lower excess free energy) as indicated by the analysis of chemical-potential stability based on density-functional theory. Therefore, such an interface likely forms during high-temperature annealing by interfacial reconstruction and atomic diffusion, even when the initial terminations are uncontrolled. DFT calculations also revealed that this interface exhibits strong polarity and chirality, despite being formed between the same material on both sides. Although the stacking order appears symmetric, the middle SrO or TiO$_2$ layer must reconstruct to conform more with the top or bottom lattice orientation, creating two possible configurations that are mirror images of each other. \replaced{This broken symmetry was experimentally confirmed by optical second-harmonic generation, where the signal from the bilayer exceeds the sum of the individual membranes. The excess arises from an interfacial polar reconstruction, whose nonlinear response can be distinguished by phase from that of the surface. With a model considering optical interference, we extracted an interfacial nonlinear susceptibility comparable in magnitude to the strong surface dipole of a bare SrTiO$_3$ surface.}{This broken symmetry was experimentally proved by optical second-harmonic generation, showing that the second-order nonlinear susceptibility of the interface is comparable to that of the bare SrTiO$_3$-air interface, which exhibits a strong surface dipole.} The emergence of the macroscopic optical properties is corroborated by atomic-scale imaging of the interface, showing a significantly thinner low-crystallinity region compared to previous reports, as well as bonding consistent with our first-principles calculations. Our multi-probe experiments and comprehensive theoretical results provide strong signature of twisted oxide interface and elucidate details about the bond formation, establishing a workflow to systematically engineer the structural and optical properties of oxide moiré superlattices.

\section{Results and Discussions}

\begin{figure}[H]
  \includegraphics[width=\linewidth]{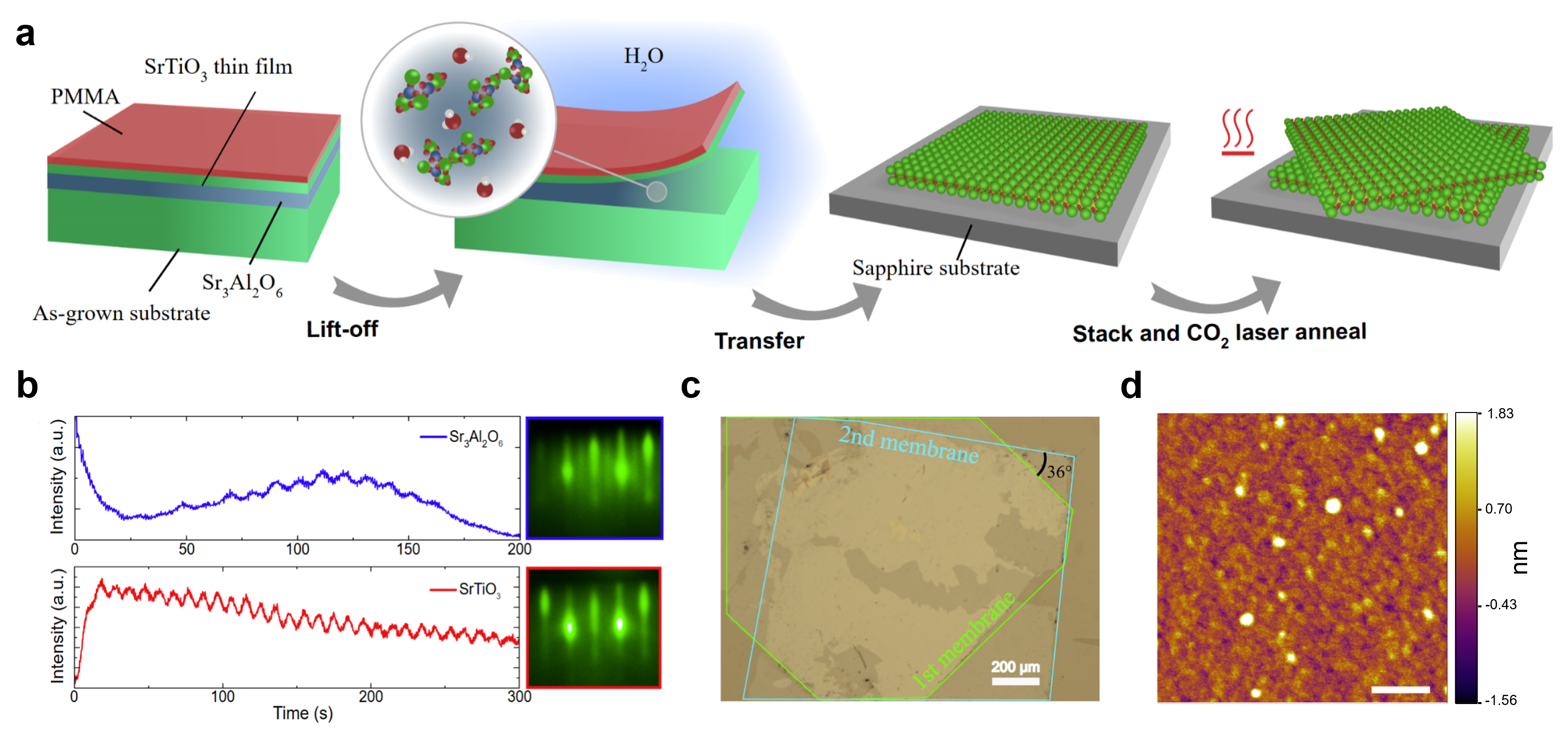}
  \caption{(a) Schematic of preparing twisted SrTiO$_3$ bilayer membranes on a sapphire substrate using thin-film epitaxy and chemical lift-off methods. (b) RHEED oscillations and patterns for Sr$_3$Al$_2$O$_6$ and SrTiO$_3$ films, respectively, indicating a 2D layer-by-layer growth mode and atomically smooth surfaces. The film thickness is directly measured from the number of unit cells. (c) Optical microscope image of the twisted SrTiO$_3$ bilayer membranes transferred on a sapphire substrate, where large, clean bilayer and monolayer regions are present. (d) Atomic force microscopy image of twisted bilayer region after annealing, showing terraces along the surface and smaller roughness. Scale bar:~\SI{1}{\micro\meter}. }
  \label{fig:fig1}
\end{figure}

We fabricated twisted SrTiO$_3$ bilayer membranes using reflection high-energy electron diffraction (RHEED)-assisted pulsed laser deposition (PLD) and subsequent chemical lift-off, transfer, and stacking processes (\textbf{\autoref{fig:fig1}a}). First, we synthesized heterostructures consisting of SrTiO$_3$ films with a thickness $t=20$ nm on (001)-oriented SrTiO$_3$ substrates with a 12-nm water-soluble Sr$_3$Al$_2$O$_6$ buffer layer serving as a sacrificial layer for lift-off. The synthesis process was monitored in situ by RHEED, where pronounced intensity oscillations and streak-like diffraction patterns indicate a 2D, layer-by-layer growth mode for both the Sr$_3$Al$_2$O$_6$ and SrTiO$_3$ layers (\autoref{fig:fig1}b). The RHEED intensity oscillations correspond to the completion of a unit-cell layer, enabling precise control of the film thickness. After synthesis, a polymethyl methacrylate (PMMA) support layer with a thickness of 600-800 nm was spin-coated onto the heterostructure to assist membrane lift-off and transfer. \deleted{The PMMA layer can be removed later using organic solvents after membrane transfer.} The heterostructure was placed in deionized water at room temperature to dissolve the sacrificial Sr$_3$Al$_2$O$_6$ layer, thereby releasing the SrTiO$_3$ film from the substrate. The released SrTiO$_3$ membrane was then transferred and laminated onto a sapphire substrate. \added{The PMMA was subsequently removed in a heated acetone bath for 30 minutes, followed by cleaning in an IPA bath to minimize organic residue, then blow-dried with nitrogen.} To ensure that the membrane conformed closely to the sapphire substrate, the SrTiO$_3$-sapphire stack was annealed at 950 $\degree$C \deleted{(Figure S1)} for one hour in a vacuum pressure of 10$^{-6}$ Torr using a resistive heating stage inside the PLD chamber. 

A second SrTiO$_3$ membrane was prepared using the same procedure and then transferred onto the annealed SrTiO$_3$-sapphire stack at a twist angle of 36$\degree$ relative to the first SrTiO$_3$ membrane. The twist angle was further confirmed by optical imaging, which reveals both the monolayer and bilayer regions across the millimeter-scale membranes (\autoref{fig:fig1}c). To achieve a clean interface between the membranes, the bilayer stack was further annealed at 1000 $\degree$C for 200 seconds under an oxygen pressure of 75 mTorr inside the PLD chamber using a CO$_2$ laser heating stage, followed by rapid cooling at the same pressure. We verified the surface flatness at the atomic scale by atomic force microscopy (AFM). Compared to the bilayer membrane stack before annealing (Figure S1), the laser-annealed twisted bilayers exhibit smoother surfaces with a root-mean-square (RMS) roughness of 345 pm, as well as a clear presence of step terraces along the diagonal direction (\autoref{fig:fig1}d). This represents a significant improvement in surface quality, likely resulting from the removal of residual organic contamination and amorphous layers.

\begin{figure}[H]
  \includegraphics[width=\linewidth]{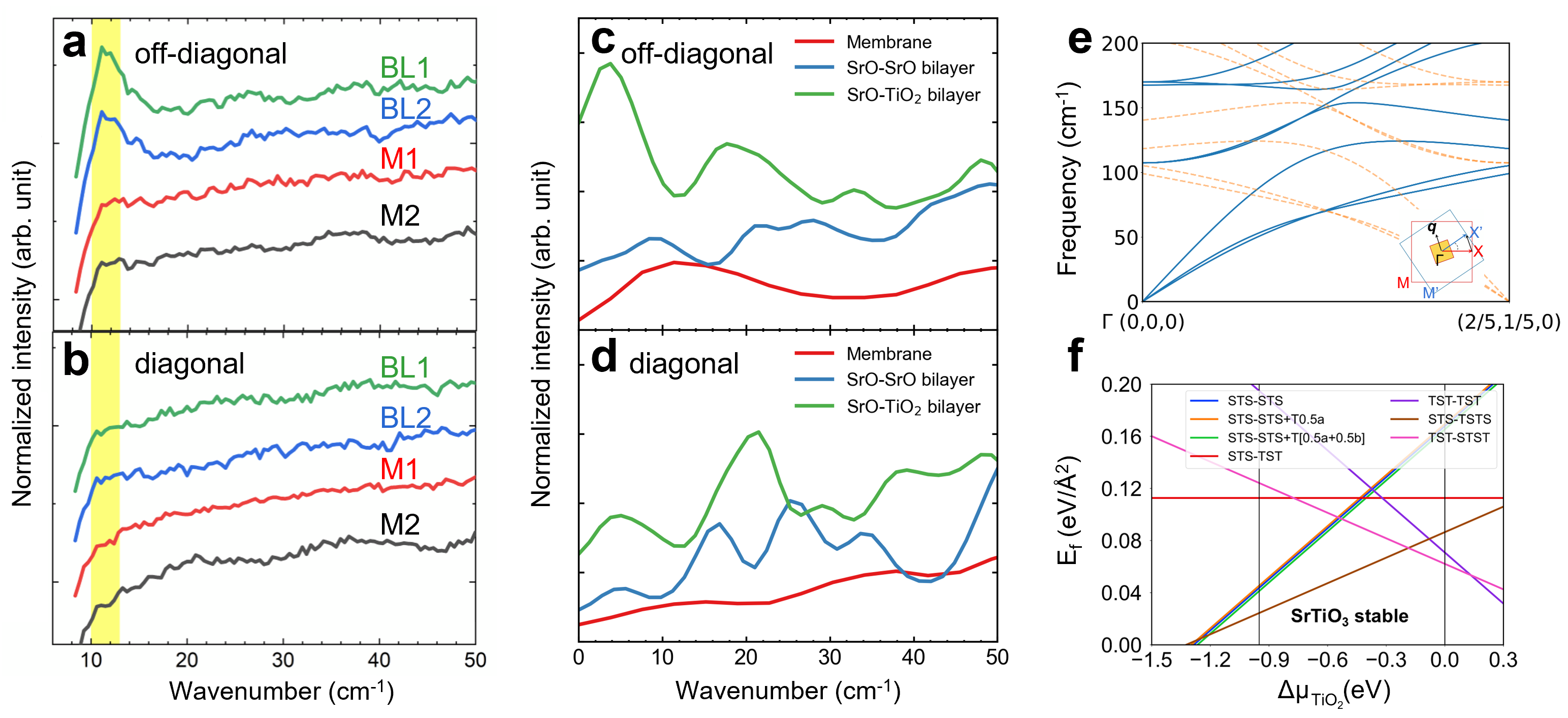}
  \caption{Low-frequency Raman modes from moiré superlattices. Helicity-resolved Raman spectra from bilayer (BL1 and BL2) and membrane (M1 and M2) areas were obtained at 150 K with (a) off-diagonal and (b) diagonal detection configuration. The new Raman feature at around 12 cm$^{-1}$ is highlighted. (c,d) Calculated Raman spectra of $\Sigma5$ interface formed by different termination types under (c) off-diagonal and (d) diagonal configurations. (e) Calculated phonon band structure of SrTiO$_3$ in cubic phase (blue solid lines), showing that all phonons with the moiré superlattice wavevector (2/5, 1/5, 0) or equivalent have higher frequencies than the observed feature. The yellow dashed lines indicate the corresponding folded phonon dispersions. Inset: Schematic of twisting-induced Brillouin zone-folding. The inset shows the original Brillouin zones of 1st (red) and 2nd (blue) membranes and the folded moiré Brillouin zone (orange). (f) The formation energy (E$_f$) of different interfacial structures as a function of chemical potential of TiO$_2$ ($\mu_{\rm {TiO_2}}$), which shows SrO-TiO$_2$ interfaces have the lowest E$_f$ in the SrTiO$_3$ stable region (determined by the bulk phase diagram in Figure S9). In the legend, T[0.5$a$]and T[0.5$a$+0.5$b$] represent translation of the upper slab along  $a$ for 0.5$a$ length and diagonally for 0.5$a$ and 0.5$b$, respectively.}
  \label{fig:fig3}
\end{figure}

We found evidence of phonons at the bilayer interface in Raman scattering. Cubic bulk SrTiO$_3$ is known to exhibit broad second-order Raman scattering\replaced{, which, although reduced, is still present in film samples~\cite{nilsen_raman_1968, perry_temperature_1967, oshea_temperature_1967,razumnaya_phase_2016,gupta2001temperature}.}{ due to the selection rules of the perovskite structure}. Fortunately, \replaced{the background is sufficiently suppressed in our membranes for observing some}{thin films suppress this broad background and make it possible to observe} surface- and interface-related, first-order, zone-center optical phonon modes as a result of a lower symmetry~\cite{sirenko_observation_1999}. Raman spectroscopy was conducted at 150 K, 50 K, and 10 K (above and below the bulk cubic-tetragonal phase transition temperature $\sim$ 105 K \cite{shirane1969lattice}), and signals were detected in both off-diagonal and diagonal circular polarization configurations (see Method section for details). Raman spectra were collected at two positions of overlapping bilayer membranes (BL1, BL2) and two positions with only a single membrane layer (M1, M2), as shown in \textbf{\autoref{fig:fig3}a-b} for the cubic phase at 150 K. In the off-diagonal channel, the bilayer regions exhibited a peak at 12 cm$^{-1}$, which is not typically observed in the Raman spectra of SrTiO$_3$. \added{The real center frequency of the mode may be lower but cannot be acquired due to the onset of filter cutoff at 11 cm$^{-1}$.} This peak was absent in membrane regions and in the diagonal channel, indicating that it originated from an interlayer interaction with a strong off-diagonal Raman tensor. The feature remains the same at 50 and 10 K, with additional Raman modes observed at 15 \deleted{, 60, }and 170 cm$^{-1}$, as shown in Figure\replaced{ S3, S4, and S5}{s S3$-$S5}. The mode at 15 cm$^{-1}$ is attributed to the bulk antiferrodistortive (AFD) soft mode activated by the tetragonal phase transition at $T = 105$~K, consistent with its strong off-diagonal Raman tensor components~\cite{fleury_soft_1968}. The peaks at 170 cm$^{-1}$ are assigned to the TO$_2$ optical phonon modes~\cite{sirenko_observation_1999,razumnaya_phase_2016,gupta2001temperature}. \added{Other possible features above 20 cm$^{-1}$ were masked by the oscillations in the broad spectra introduced by the interference from the cryostat window in the system (\replaced{Figure}{Fig.} S3).} \deleted{The mode near 60 cm$^{-1}$ can originate from two vibrational modes close to each other: the out-of-plane Ti/Sr distortion mode and the interlayer breathing mode (Figure S14), both brightened in Raman by AFD. Clearly,} Since only the 12 cm$^{-1}$ peak is insensitive to phase transition \replaced{and does not appear in single-layer membranes, this feature can}{, it is }not be related to the bulk symmetry of SrTiO$_3$ but must come from interfacial reconstruction.

Two potential mechanisms could account for this peak: (i) moiré potential-induced folding/scattering of bulk phonon bands and (ii) interfacial vibrational mode arising from the twisted stacking. Given the periodicity of the moiré superlattice, bulk acoustic phonons away from the center of the Brillouin zone could be folded onto the $\Gamma$ point and potentially give rise to new Raman activity. However, under our coincidence site lattice condition, the reciprocal lattice vector of the moiré superlattice is quite large and also commensurate with that of the bulk SrTiO$_3$. Hence, the estimated frequency for such folded modes, even for the lowest acoustic branches, is approximately 100 cm$^{-1}$ for the cubic structure~\cite{tadano_ab_2019} as shown in \autoref{fig:fig3}e at the moiré reciprocal lattice vector (2/5, 1/5, 0), which does not explain the 12 cm$^{-1}$ peak. Consequently, the moiré folding of bulk modes was ruled out \added{as an explanation}. 

We attribute this peak to an interfacial shear mode activated by polar bonding according to \textit{ab initio} molecular dynamics (AIMD). First, we chose twisted bilayer structures with different terminations and determined the energy stability of each structure (\autoref{fig:fig3}f). Here S and T represent a single SrO layer and TiO$_2$ layer, respectively. In order to determine the stability of the interface structures, we first determined the chemical potential ranges ($\Delta \mu \rm_{Ti{O_2}}$) where SrTiO$_3$ becomes dominant over other strontium titanates through bulk phase diagram by first-principles thermodynamic calculations (Figure S9).\cite{reuter2003first, qin2024unveiling} Then by performing first-principles calculations of the formation energy (E$_f$, \replaced{see Method section for details}{see methods section}) as a function of $\Delta \mu \rm_{Ti{O_2}}$, we found that the SrO-TiO$_2$ interface (STS-TSTS and TST-STST) is energetically favored over the SrO-SrO and TiO$_2$-TiO$_2$ interface in the bulk stability region of SrTiO$_3$. Additionally, we noticed that the interfacial TiO$_2$ layer tends to bind more strongly to one side of the slabs than to the other, forming a globally chiral moiré superlattice that lacks mirror and inversion symmetry. The twist between the slabs provides a built-in mechanical field that drives chiral deformation of the interfacial region, a phenomenon that could be termed chiral mechanochemistry. \cite{do2017mechanochemistry,liu2022review, cheng2022assembly, zhou2022chiral}. 

We then calculated the time-domain autocorrelation function of polarizability for the low-energy structures and used the Fourier transform to obtain Raman spectra~\cite{zhang2023discovery,yaffe2017local}. Three distinct models were studied: an SrTiO$_3$ membrane (red), a twisted bilayer with an SrO-SrO interface (blue), and a twisted bilayer with an SrO-TiO$_2$ interface (green). The simulated Raman spectra for both diagonal and off-diagonal configurations\added{ at 150 K} are plotted in \autoref{fig:fig3}c-d (see Method \replaced{section}{Section} for details). Notably, \replaced{only}{both} the \deleted{SrO-SrO and }SrO-TiO$_2$ interface exhibits a new distinct peak at approximately 5--10 cm$^{-1}$ \added{in the off-diagonal channel} compared to the SrTiO$_3$ membrane\replaced{ and SrO-SrO interface. The accuracy of the center frequency is limited by the finite time window of AIMD and polarizability correlation for such a large system.}{, with the same selection rule observed in our experiment. This suggests that superlattice bonding at the twisted interface gives rise to a new vibrational mode.} To \replaced{understand the new}{elucidate the} vibrational mode, we performed a frequency-filtered analysis for the atomic displacement trajectories, which indicates that the peak is associated with interlayer in-plane shearing displacement in the bilayer structure, as shown in Figure\deleted{s} S14, S15, S16, and Supporting Videos 1--4. \replaced{This shearing mode is almost doubly degenerate, as the atoms move in both x and y directions at different times, leading to the observed Raman selection rule in the off-diagonal channel. By contrast, although the SrO-SrO interface also exhibits an interlayer shearing mode, it has a much higher frequency and mainly displays Raman activity in the diagonal channel.}{Interestingly, the SrO-TiO$_2$ interface shows much more pronounced Raman activity, indicating more broken symmetry at the interface than the SrO-SrO and SrO-TiO$_2$ interfaces.} \added{We also compared the calculated monolayer STO, which showed no Raman activity at 150 K, as expected.} \replaced{Overall, the new low-frequency, off-diagonal mode indicates a primarily}{Thus, the much larger strength of the observed new peak in the bilayer interface compared with the surface modes in the tetragonal phase could be an indication of the} SrO-TiO$_2$ interface\added{ in our bilayer samples}.

\begin{figure}[H]
  \includegraphics[width=\linewidth]{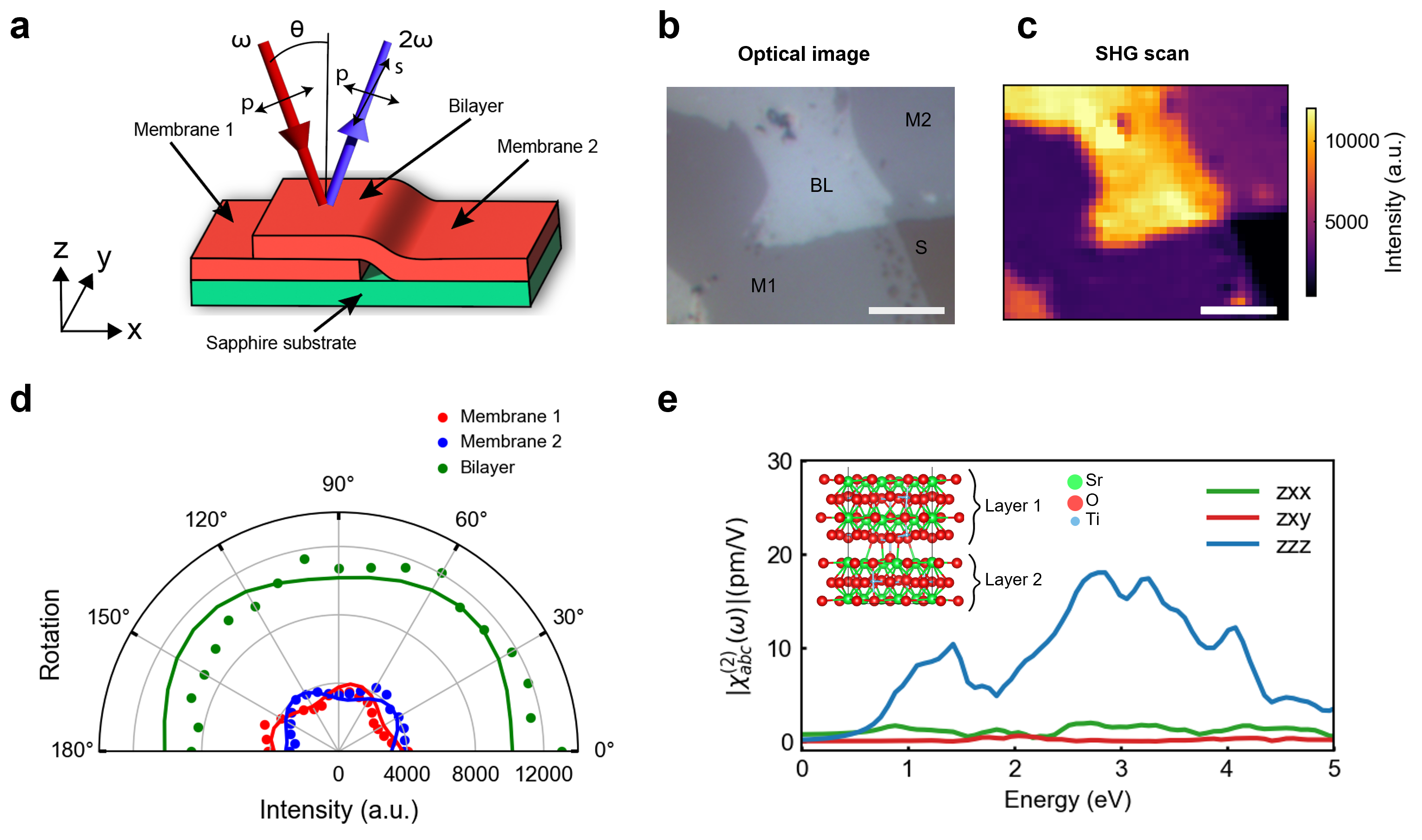}
  \caption{(a) Schematic of the angle-dependent scanning SHG microscopy geometry in reflection mode. (b) The optical microscopy image distinguishes one membrane region (M1), the other membrane region (M2), the bilayer overlapping region (BL), and the bare sapphire substrate (S). (c) Reflective scanning SHG image of the same area showing enhanced SHG signal in the bilayer, compared with the top (M1) and bottom (M2) membranes. Scale bar: \SI{10}{\micro\meter}. (d) Surface SHG of SrTiO$_3$ as a function of sample orientation measured in reflection geometry with p-polarization excitation and polarization-unresolved detection (Figure S6\replaced{ and }{-}S7). The solid curve shows the model fitting from \replaced{Equation}{Eqs.}~\ref{eqn:Efields} and \ref{eqn:intensities}, accounting for interference between the top and bottom surfaces. (e) DFT-calculated all non-zero $\chi^{(2)}$ tensor values of SrO-TiO$_2$ interface. Note: $\chi_{zxx}$ is shown; $\chi_{zyy}$ and $\chi_{xxz}$ are symmetry-equivalent.}
  \label{fig:fig4}
\end{figure}

To further verify the symmetry-breaking of interfacial bonding and understand its effect on the structural and electronic states of the bilayer, we performed angle-dependent second-harmonic generation (SHG) imaging in reflectance configurations. To emphasize the surface and interfacial effect, we ensured a finite $E_z$ component at an incidence angle of $20\degree$ in p-polarization (\textbf{\autoref{fig:fig4}a}). The detection polarization was not selected for the rotational SHG measurement to capture contributions from all polarization components. We observed a clear contrast of SHG between the two single-membrane regions (M1, M2) and the bilayer membrane (BL) (\autoref{fig:fig4}b-c), and the SHG of the bilayer ($I_{BL}$) is stronger than the summation of the single-layer regions ($I_{M1}$+$I_{M2}$) (\autoref{fig:fig4}d). SHG signals are reported in arbitrary units because the absolute collection efficiency was not calibrated. By comparing the excitation power needed to achieve the same SHG intensity from monolayer semiconductors, which is two orders of magnitude smaller, we estimated that the effective nonlinear susceptibility is on the order of 2 pm/V assuming an effective thickness of 1 nm~\cite{biswas_non-linear_2023}. A naive comparison might suggest that the introduction of the additional SrTiO$_3$-SrTiO$_3$ bilayer interface (I) adds a finite contribution to the measured SHG intensity. However, quantitative attribution requires accounting for optical interference effects in the multilayer stack to isolate the interfacial nonlinear response.

SHG mainly arises from two mechanisms: (i) surface dipole contributions, which are allowed due to the broken inversion symmetry at interfaces, and (ii) bulk electric-quadrupole or magnetic-dipole contributions, which are higher-order in nature~\cite{sipe_phenomenological_1987}. In our SrTiO$_3$ membrane bilayer supported on sapphire, the total SHG response is therefore expected to be a superposition of three terms: (i) the dipolar response from the SrTiO$_3$ surfaces with $4mm$ symmetry~\cite{zhao_study_2013}, $E_{i}=\chi_{ijz,ML}^{(2)} E_j E_z$, and that from the sapphire surface with 6mm symmetry, $E_{i}=\chi_{ijz,S}^{(2)} E_j E_z$, (ii) the bulk electric quadrupolar contribution from SrTiO$_3$ crystal (denoted as B, with $m\overline{3}$m symmetry) and from sapphire (S, with $\overline{3}$m symmetry), $E_{i}=\chi_{ijkl,B/S}^{(3)} \nabla_j E_k E_l$, and (iii) the interface dipolar contribution from the twisted SrTiO$_3$-SrTiO$_3$ bilayer, $E_{i}=\chi_{ijz,H}^{(2)} E_j E_z$. Here, $i$, $j$, $k$, and $l$ are labels of coordinate axes in the laboratory frame. The symmetry-allowed terms yield the following form of second\added{-}harmonic fields as a function of sample orientation ($\theta$) for each part of the stack. Because SHG emission propagates backward against incident light, there is a significant phase delay $\phi=2\pi(n_{800 nm}+n_{400 nm})t/(400~\mathrm{nm})=1.55$~rad between the top and bottom surfaces of a single membrane, calculated from the refractive indices $n$ at the pump and detection wavelengths and the thickness of the membrane. For mathematical symmetry, we set the phase at the interface to zero and note that the top (t) and bottom (b) surfaces of the membranes should have exactly opposite $\chi^{(2)}$. Finally, we verified that bare sapphire at any incidence angle and SrTiO$_3$ single crystal at normal incidence give negligible SHG signals, so these terms were not included. 

\begin{equation} \label{eqn:Efields}
\begin{alignedat}{1}
&E_{M1,\text{t}} = \left( A_{11} + i A_{12} \right) e^{-i \phi}, \\[-2pt]
&E_{M1,\text{B}} = iA_{21} \cos\!\big[ 4(\theta + \delta_{1}) \big] e^{-i \phi / 2}, \\[-2pt]
&E_{M1,\text{b}} = - \left( A_{11} + i A_{12} \right), \\[-2pt]
&E_{I} = A_{31}+iA_{32}, \\[-2pt]
&E_{M2,\text{t}}  = \left( A_{11} + i A_{12} \right), \\[-2pt]
&E_{M2,\text{B}} = iA_{21} \cos\!\big[ 4(\theta + \delta_{2}) \big] e^{i \phi / 2}, \\[-2pt]
&E_{M2,\text{b}} = - \left( A_{11} + i A_{12} \right) e^{i \phi}.
\end{alignedat}
\end{equation}
All fitting coefficients contain real and imaginary parts accounting for dispersion and absorption, but we can set the relative phase of $A_{21}$ (quadrupole contribution of the bulk SrTiO$_3$) to zero without affecting the total intensity. The phase is likely small because SrTiO$_3$ is transparent in both the fundamental and harmonic frequencies, but the surface and interface can host lossy states. The angles of lattice orientation $\delta_1$ and $\delta_2$ account for the azimuthal alignment of the membranes. The total SHG intensities of membrane 1, 2, and the bilayer can be expressed as:
\begin{equation} \label{eqn:intensities}
\begin{alignedat}{2}
&I_{M1} = \left| E_{M1,t} + E_{M1,B} + E_{M1,b} \right|^{2}, \\[-2pt]
&I_{M2} = \left| E_{M2,t} + E_{M2,B} + E_{M2,b} \right|^{2}, \\[-2pt]
&I_{BL} = \left| E_{M1,t} + E_{M1,B} + E_{I} + E_{M2,B} + E_{M2,b} \right|^{2}.
\end{alignedat}
\end{equation}

After simplification (Supporting Information, Section S4), $I(\theta)$ exhibits 4-fold symmetry, as expected, and contains components independent of angle primarily coming from the surface terms, as well as a $\cos(4\theta)$ component from the bulk-surface interference~\cite{sipe_phenomenological_1987}. By Fourier transforming the
experimental data $I_{M1}$ and $I_{M2}$, we obtained the best fitting parameters: $A_{11} = 42.4,  A_{12} = 0.3, A_{21} = -4.2$. The difference between the fitted $\delta_1$ and $\delta_2$ is very close to the expected twist angle $36\degree$. Using the parameters for M1 and M2, we then fitted $I_{BL}$ and obtained $A_{31} = 24.7, A_{32} = -14.9$, that is, $|A_{31}+iA_{32}|/|A_{11}+iA_{12}|\approx0.68$, indicating a strong interfacial nonlinearity comparable to surfaces, which is surprising given that the two sides are the same material and should have the same electrostatic potential without significant polarization. 

The large polarization can be understood in terms of the relaxed interfacial structure and the nonlinear susceptibility calculated by DFT. If a twisted interface is formed by symmetric terminations (SrO-SrO or TiO$_2$-TiO$_2$), there would only be a small SHG signal arising from structural chirality (Figure S13). Instead, for an interface with asymmetric terminations (SrO-TiO$_2$), the SHG signal is much larger in all components, as shown in \autoref{fig:fig4}e. To further clarify the microscopic origin, we systematically varied four factors in our calculations--the terminations at the bonded interface, the termination of the free-surface, the local stacking geometry, and the number of layers--and found that the enhancement of the SHG signal is overwhelmingly governed by the terminations at the bonded interface, whereas the other factors lead to comparatively minor variations. Consistent with this picture, our DFT electronic structure of the moiré bilayer shows flat-band-like dispersions and an enhanced density of states near the band edges (Figure \replaced{S11}{S14}), which increases the number of optically accessible channels and can strongly amplify the nonlinear response. Such near-resonant effects also agree with the experimentally observed finite imaginary part of the interfacial SHG. The dominant component of the interfacial $\chi^{(2)}$ in such a reconstruction of the symmetry group $4$ is $zzz$, which gives an angle-independent SHG signal in the p-p channel, also in agreement with what we observed in the experiment, as opposed to the signal expected in the p-s channel for a pure chirality-induced $zxy$ component from $422$ symmetry (Figure S8). Although we cannot control membrane termination during fabrication (i.e., SrO-TiO$_2$ vs TiO$_2$-SrO), the SHG from a random distribution of terminations will not cancel completely and statistically has a finite value (Supporting Information, Section S5)~\cite{biswas_non-linear_2023}.

\begin{figure}[H]
  \centering
  \includegraphics[width=0.9\linewidth]{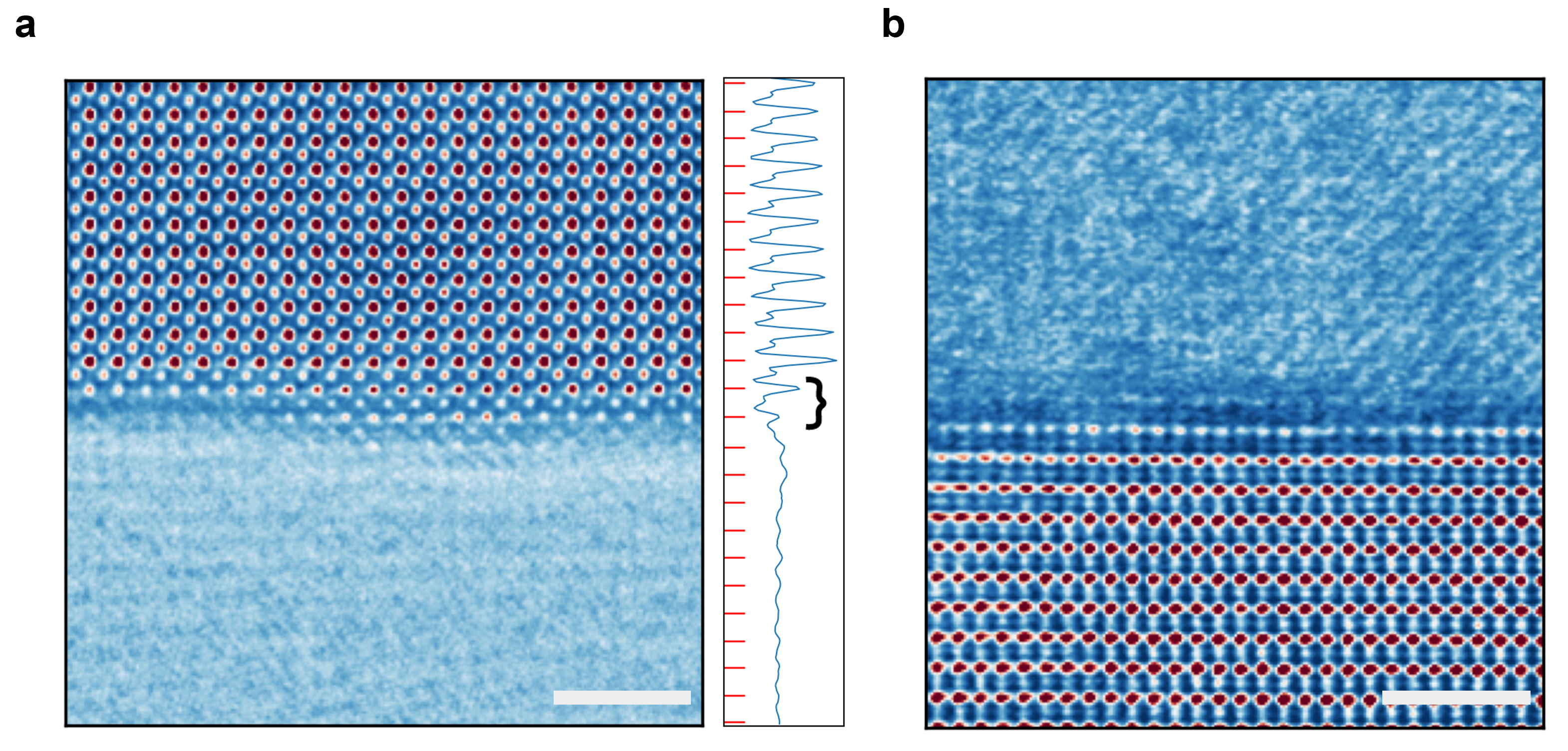}
  \caption{Cross-sectional HAADF-STEM image of the twisted post-annealed membrane stack. (a) Images acquired along the [100] and (b) [110] zone axes of the atomically resolved crystal, respectively. In each case, the imaged membrane is aligned on-axis, while the opposing membrane is oriented off-zone axis. The line-integrated intensity is shown on the right side of each image. Red lines indicate the peak intensities distanced by one unit cell at the bulk and 10\% more at the interface, consistent with the calculated increase in SrO-SrO distance at the twisted interface without notable van der Waals gaps. The brace indicates the transition region, characterized by roughness and/or reduced crystallinity.
  Scale bar: \SI{2}{\nano\meter}.}
  \label{fig:fig2}
\end{figure}

Lastly, we confirmed the formation of a clean twisted SrTiO$_3$ interface that is consistent with the optically observed interlayer interactions from cross-sectional scanning transmission electron microscopy (STEM) imaging (\replaced{see Method section for details}{Methods}). The high-angle annular dark-field (HAADF) STEM image viewed along the [100] zone axis of the top layer SrTiO$_3$ reveals high crystalline quality of the thin film (\textbf{\autoref{fig:fig2}a-b} and Figure S2). Due to the 36$\degree$ twist angle, the lattice of the bottom layer appears blurred because its zone axis is misaligned with the projection direction. The constituent bilayer SrTiO$_3$ membranes form an intimate interface without notable van der Waals gaps, as shown by the spacing between the intensity peaks of the Sr atomic column extracted from the integrated line profile (\autoref{fig:fig2}a). The periodicity evidences that an alternating SrO-TiO$_2$ interface is more favorable. The peak distance at the interface is 10\% larger than the unit cell length, because the theoretically predicted gap between the twisted SrO layers with TiO$_2$ bonds in between is wider (Figure S15). Although the interface is not atomically sharp, its roughness closely matches that of the top surface, with height variations of approximately one unit-cell thickness. The interfacial region exhibits lower scattering intensities, which may result from roughness along the viewing direction, lower crystallinity, or partial bonding. Imaging along the zone axis of the bottom membrane also shows intimate lattice contact with the top membrane, with a transition region of two SrO layers. In certain locations, the transition region is slightly thicker (\replaced{Supporting Information, Section}{SI section} S2) but no more than those reported in previous studies~\cite{Sanchez-Santolino2024-gy, Kim2025-lk, Zhang2025-iv}. This transition region represents the interfacial reconstruction at the twisted interface to accommodate interfacial strain and chemical bonding. Given that the bottom membrane is not perfectly terminated, some spatial variation in transition-layer contrast is expected following interfacial bonding. The atomic bonding observed at portions of the interface is consistent with our first-principles calculations and accounts for the emergence of the Raman shear mode and second-order nonlinearity.

\section{\added{Conclusion}}
In conclusion, we experimentally demonstrated chemically bonded oxide interfaces that exhibit new Raman-active shear modes at low wavenumbers and electrical polarization with strong second-harmonic generation. These symmetry-breaking phenomena occur when the bulk is in a cubic phase without the need for lattice distortion or strain. Using systematic first-principles calculations, including SHG and AIMD-based Raman spectrum calculations, we found that these observations are most consistent with a twisted interface between asymmetric terminations spontaneously formed by annealing at high temperature. Our results provide guidance for achieving specific lattice reconstructions in moiré oxide superlattices at large twisting angles and small periodicity, and pave the way for strongly modulating the electronic moiré band structure with large Coulomb interactions towards strongly correlated phases at high temperatures. 


\section{Methods}
\subsection{Thin film synthesis}
Epitaxial heterostructures comprised of 20 nm SrTiO$_3$ films with 12 nm Sr$_3$Al$_2$O$_6$ sacrificial layers were synthesized on (001)-oriented, as-grown SrTiO$_3$ substrate by a NBM Design RHEED-assisted pulsed laser deposition system. The Sr$_3$Al$_2$O$_6$ sacrificial layers were synthesized in vacuum at a pressure of 10$^{-6}$ Torr, a heater temperature of 950 $\degree$C, a laser fluence of 2.0 J/cm$^{2}$, a spot size of 2.6 mm$^{2}$, and a laser repetition rate of 1 Hz. The SrTiO$_3$ films were synthesized at a dynamic oxygen pressure of 0.1 Torr, a heater temperature of 760 °C, a laser fluence of 1.23 J/cm$^{2}$, a spot size of 2.6 mm$^{2}$, and a laser repetition rate of 1 Hz. RHEED oscillations and diffraction patterns were recorded using a STAIB Instruments TorrRHEED system at an electron energy of 20 kV and analyzed with ksa 400 software from k-Space Associates.
\subsection{Optical second\added{-}harmonic generation}
The excitation source consisted of a Light Conversion FLINT femtosecond oscillator driving an APE Levante IR optical parametric oscillator (OPO), followed by the APE HarmoniXX SHG/THG module. The laser system produced pulses with a duration of approximately \SI{120}{\femto\second} at a repetition rate of \SI{76}{\mega\hertz} and a fundamental wavelength of \SI{800}{\nano\meter}. The beam was focused onto the sample using a 40× objective (f = \SI{5}{\milli\meter}),  with a beam diameter of \SI{2.5}{\milli\meter} (NA$_{\text{eff}}$ = 0.25), yielding a spot size of approximately ~\SI{2}{\micro\meter}. The incident power on the sample was maintained at 100 mW at a 20$\degree$ angle of incidence, achieved by laterally offsetting the beam at the back aperture of the objective lens by \SI{1.7}{\milli\meter}. In the home-built reflection-geometry SHG setup, the sample was mounted on a rotational stage positioned after the objective for measurements of rotational anisotropy. The stage was integrated with a motorized scanning platform to enable spatially resolved SHG imaging. The reflected SHG signal was filtered using a 785 nm short-pass filter and a 400 nm band-pass filter, and analyzed using a thin-film polarizer. The emitted signal was collected and detected with a single-pixel photon counter (Hamamatsu C11202-100).
\subsection{Scanning transmission electron microscopy}
The cross-sectional twisted SrTiO$_3$ sample was prepared using a FEI Helios 660 focused ion beam (FIB). After sample preparation, high-angle dark-field scanning transmission electron microscopy (HAADF-STEM) imaging was performed using an FEI Titan Themis G3 (scanning) transmission electron microscope equipped with double Cs-correctors operated at 300 kV. The convergence semi-angle is 25 mrad.
\subsection{Raman spectroscopy}
Raman measurements were performed using a Renishaw inVia confocal microscope system equipped with a Montana Instruments closed-cycle cryostat. The sample was excited with a 532 nm CW laser at a power of 2 mW, focused through a 100× objective (NA = 0.9). To control the polarizations of incident and collected light, quarter-wave plates and a linear polarizer were employed. For high-resolution, low-wavenumber measurements, a 3000 g/mm grating and a notch filter were utilized. The diagonal channel is defined as the circularly polarized electric field that rotates in the same direction for both excitation and scattered light. In transmission geometry, this is denoted as the co-polarized configuration (often labeled the left-left or right-right channel in literature). But in reflection geometry, strictly speaking, the emission polarization is opposite to the excitation polarization. Similarly, the off-diagonal channel is defined as the circularly polarized electric field rotating in the opposite direction for excitation and scattered light.

\subsection{First-principles calculation}
We performed plane-wave DFT calculations using the \textsc{Quantum ESPRESSO} package \cite{espresso_giannozzi2009quantum,espresso_giannozzi2017advanced} with optimized norm-conserved pseudopotential (OPIUM) and Perdew-Burke-Ernzerhof (PBE) exchange-correlation functional.\cite{rappe1990optimized,ramer1999designed,perdew1996generalized} Grimme's D3 dispersion correction was applied for all calculations to consider the Van der Waals interactions.\cite{grimme2006semiempirical,grimme2010consistent} For the modeling of the slabs, a vacuum region of at least 20~\AA\ was introduced to avoid spurious interactions between periodic images. For structural relaxation, a cutoff energy of 80 Ry and a $6\times6\times1$ $k$-point mesh were used. The convergence criteria for total energy and atomic forces were set to 1 × 10$^{-6}$ Ry/unit cell and 1 × 10$^{-3}$ Ry/Bohr, respectively. For \textit{ab initio} molecular dynamics (AIMD), we employed \added{$2\times2\times1$ $k$-point mesh for the membrane model} and $\Gamma$-point-only sampling \added{for the bilayer structures} with a total-energy convergence threshold of $1\times10^{-6}$ Ry/unit cell. Dielectric-constant calculations were carried out using a $2\times2\times1$ $k$-point mesh \added{for every 100 fs to obtain the polarizability}. The three atomic structures were \replaced{considered}{considerd} in AIMD: the membrane consisting of three atomic layers (a supercell of $(2,0,0),(0,2,0)$ with a total of 28 atoms is used to match the lateral dimension of the bilayer structures), the SrO-SrO bilayer consisting of six atomic layers with a total of 70 atoms, and the SrO-TiO$_2$ bilayer consisting of seven atomic layers with a total of 85 atoms. For the nonlinear optical calculations, we used a $6\times6\times1$ $k$-point mesh for the self-consistent step and a denser $30\times30\times1$ mesh for the non-self-consistent step, based on a $\Sigma5$ supercell with seven layers and stacking sequences, the SrO-TiO$_2$ interface bilayer contains seven atomic layers; a tighter total-energy convergence threshold of $1\times10^{-10}$ Ry/unit cell was used.

\subsubsection{Raman-MD}
AIMD was performed on a SrTiO$_3$ membrane, a twisted SrTiO$_3$ bilayer with SrO-SrO interface, and a twisted SrTiO$_3$ bilayer with SrO-TiO$_2$ interface at 30 K using the Born-Oppenheimer MD scheme with 1 femtosecond (fs) per step. The ionic temperature is controlled by a velocity rescaling method, with rescaling every 50 fs to the target temperature. A total of about 25 picoseconds (ps) trajectories were obtained for each structure, and the first few picoseconds were abandoned for temperature stabilization. To obtain the Raman spectra, we calculated the polarizability tensor for every 100-fs snapshot. In linear response, the dielectric tensor $\boldsymbol{\epsilon}$ and the polarizability tensor $\boldsymbol{\alpha}$ are related by
\begin{equation} \label{eqn:dielectric}
\begin{alignedat}{3}
& \boldsymbol{\epsilon} = \boldsymbol{I} + \frac{1}{\epsilon_0 V}\boldsymbol{\alpha}
\end{alignedat}
\end{equation}
where $\boldsymbol{I}$ is the identity matrix, $\epsilon_0$ is the vacuum permittivity, and $V$ is the volume of the unit cell.
The Raman tensor component is the Fourier transform of the autocorrelation function of the polarizability component :
\begin{equation} \label{eqn:raman_tensor}
\begin{alignedat}{4}
& R_{ij}(\omega) = \int{<\alpha_{ij}(\tau)\alpha_{ij}(t+\tau)>e^{-i\omega t} dt}
\end{alignedat}
\end{equation}
where $\omega$ is the Raman spectrum frequency, $\alpha_{ij}$ is the $ij$ component of the polarizability tensor, $t$ is the time and $\tau$ is the integration time lag.

\added{In a circular polarization Raman spectrum, we define}:
\begin{equation} \label{eqn:intensity}
\begin{alignedat}{5}
& \alpha_{iso} = \alpha_{xx} + \alpha_{yy},\ \alpha_{aniso} = (\alpha_{xx}-\alpha_{yy}) + 2i\alpha_{xy} \\
& R_{iso} = \int{<\alpha_{iso}(\tau)\alpha_{iso}(t+\tau)>e^{-i\omega t} dt},\ R_{aniso} = \int{<\alpha_{aniso}(\tau)\alpha_{aniso}(t+\tau)>e^{-i\omega t} dt} \\
& I_{\sigma+\sigma-} = \frac{1}{4}R_{iso}\frac{\omega}{1-exp(-\frac{\hbar\omega}{k_B T})},\ I_{\sigma+\sigma+} = \frac{1}{4}R_{aniso} \frac{\omega}{1-exp(-\frac{\hbar\omega}{k_B T})}
\end{alignedat}
\end{equation}
where $\hbar$ is reduced Planck constant, $k_B$ is the Boltzmann constant, and $T$ is the temperature.

\subsubsection{Thermodynamic stability}
To evaluate the thermodynamic stability of the various interface structures, the formation energy (E$_f$) of each possible interface configuration with respect to the chemical potential of TiO$_2$ ($\Delta \mu_{\rm TiO_2}$) is calculated by
\begin{equation}
    E_{f} =\frac{1}{A}(E_{I} - n_{\mathrm{Sr}}E_{\mathrm{SrTiO_3}}- \left( n_{\mathrm{Ti}} - n_{\mathrm{Sr}} \right) (E_{\mathrm{TiO_2}}+\Delta \mu_{\mathrm{TiO_2}}))
\end{equation}
where E$_I$ is the total energy of the interface structure, while E$\rm _{SrTiO_3}$ and E$\rm _{TiO_2}$ represent the total energy of the bulk SrTiO$\rm _3$ and TiO$\rm _2$, respectively. $A$ is the surface area, and n$_i$ is the number of atoms in the interface for element $i$. The stability region of SrTiO$\rm_3$ in terms of $\Delta \mu_{\rm TiO_2}$ is determined by the bulk stability diagram in Figure S9, and the most stable twisted SrTiO$_3$ interface should be chosen under the bulk stability region of SrTiO$\rm_3$.\cite{reuter2003first, qin2024unveiling} 

\subsubsection{Nonlinear optical response}
The second-order nonlinear susceptibility tensor $\chi^{(2)}$ was computed from first-principles electronic structures within the independent-particle approximation using a length-gauge formulation of the second-order optical response\cite{aversa_sipe_1995,sipe_shkrebtii_2000}. All atomic structures (monolayer, bilayer, and interfacial models with different terminations) were fully relaxed until the residual forces and total-energy changes were below the chosen thresholds. A slab/supercell geometry was employed with sufficient vacuum along the out-of-plane direction to avoid spurious interactions between periodic images. Convergence was carefully checked with respect to $k$-point sampling, the number of unoccupied bands included in the summations, and the vacuum thickness. For comparison among interfaces with different terminations, we report all symmetry-allowed nonzero tensor components in the crystallographic frame and, when needed, transform them to the laboratory frame used in the optical geometry.

\section*{\replaced{Acknowledgements}{Acknowledgment}}
The authors thank Prof. Ramamoorthy Ramesh and Prof. Eugene J. Mele for helpful discussions. T.A.M.R.S\added{.} and H.Z. acknowledge support from the National Science Foundation (NSF) under grant number DMR-2240106\added{ and Welch Foundation (C-2311)}. X.L. acknowledges support from the Rice Advanced Materials Institute (RAMI) at Rice University as a RAMI Postdoctoral Fellow. X.L. and Y.H. acknowledge support from NSF (FUSE-2329111 and CMMI-2239545) and Welch Foundation (C-2065). X.L. and Y.H. also acknowledge the Electron Microscopy Center, Rice University. K. K. acknowledges the support from the Army Research Office under award No. W911NF-25-1-0201. R.X. acknowledges the support from the National Science Foundation (NSF) under award No. DMR-2442399. F.M. and S.H. acknowledge the support from NSF (FUSE-2329111, ECCS-2246564, and ECCS-1943895) and Welch Foundation (C-2144). B.K. and A.M.R. acknowledge support from the U.S. Department of Energy, Office of Science, Basic Energy Sciences, under Award No. DE-SC0026196 for analysis of moiré band structure and optical spectroscopy. X.H., S.Q., and A.M.R. acknowledge support by the Office of Naval Research, under grant number N00014-24-1-2500 for the study of bonding, composition, and structural dynamics in SrTiO$_3$ interfaces. Computational support was provided by the National Energy Research Scientific Computing Center (NERSC), a U.S. Department of Energy, Office of Science User Facility located at Lawrence Berkeley National Laboratory, operated under Contract No. DE-AC02-05CH11231.

\section*{Data Availability Statement}
\added{The data that support the findings of this study are available in the Rice Research Repository.}

\bibliographystyle{MSP}
\bibliography{main}

\section*{Supporting Information}
\added{Supporting Information is available from the Wiley Online Library or from the author.}

\end{document}